\documentclass[11pt]{article} 
\usepackage{latexsym}
\usepackage{amssymb}
\usepackage{amsmath}
\usepackage{hyperref}
\hypersetup{colorlinks=true, allcolors=blue}
\usepackage{graphicx}

\usepackage{slashed}
\usepackage{mathbbol}

\newcommand{\bridge}{Bridge\nolinebreak\hspace{-.05em}\raisebox{.4ex}{\tiny\bf +}\nolinebreak\hspace{-.10em}\raisebox{.4ex}{\tiny\bf +}}

\usepackage{color}

\usepackage{cite}

\begin{document}
\pagestyle{empty}

\begin{flushright}
KEK-TH-2667
\end{flushright}

\vspace{3cm}

\begin{center}

{\bf\LARGE
QED 5-loop on the lattice
}
\\

\vspace*{1.5cm}
{\large 
Ryuichiro Kitano
} \\
\vspace*{0.5cm}

{\it 
KEK Theory Center, Tsukuba 305-0801,
Japan\\
Graduate University for Advanced Studies (Sokendai), Tsukuba
305-0801, Japan\\
}

\end{center}

\vspace*{1.0cm}

\begin{abstract}
{\normalsize
We report the result of the numerical lattice computation of the
lepton anomalous magnetic moment in QED up to five loops. We
concentrate on the contributions from diagrams without lepton loops,
which are the most difficult part of the calculation in the Feynman
diagram method while the lattice formulation is the easiest.
Good agreement with the results of the Feynman diagram method is
observed.
}
\end{abstract} 

\newpage
\baselineskip=18pt
\setcounter{page}{2}
\pagestyle{plain}
\baselineskip=18pt
\pagestyle{plain}

\setcounter{footnote}{0}


\section{Introduction}
The anomalous magnetic moment of the electron and the muon,
$(g-2)_{e/\mu}$, are the most well measured and most precisely
calculated quantities in particle physics.
The comparison between the theory calculations and the experimental
measurements has given us the precision test of the quantum field
theory at the level of $10^{-10}$.
The largest contribution to $g-2$ is that from QED, where the
coefficients of the perturbative expansion in terms of $\alpha \simeq
1/137$ are calculated up to the order of $\alpha^5$, corresponding to
the Feynman diagrams which contain five loops.
There are more than 10,000 diagrams at the five-loop order, and the
estimation of the size of those contributions has been necessary for
the comparison between the theory and experiments. (For reviews, see
Refs.~\cite{Aoyama:2019ryr, Aoyama:2020ynm}.)

Two independent groups have achieved the complete calculation of the
five-loop coefficients numerically~\cite{Aoyama:2012wk,
Volkov:2024yzc}. The most difficult part of the calculation is the
contributions from the diagrams without lepton loops as such diagrams
give the most severe infrared (IR) divergences. There are about 6,000
diagrams of such type and somehow give the dominant part of the total
contributions.
Until recently, there have been a discrepancy in the no-lepton-loop
part of the calculation between two groups. The resolution of the
discrepancy has been reported in a recent
workshop~\cite{g-2_theory_initiative2024}.

In this paper, we report the results of the independent method of the
calculation based on the lattice QED simulation.
The method to evaluate $g-2$ on the lattice has been proposed and
developed in Refs.~\cite{Kitano:2021ecc, Kitano:2022gzy}. In
particular, the no-lepton-loop part of the calculation corresponds to
ignoring the fermion determinant in the path integral. In QED,
ignoring the fermion determinant leads to the free theory of the gauge
field, $A_\mu$, and thus the path integral can be trivially performed
by generating gaussian noises. 
Also, the absence of the lepton loop means that there is no
renormalization of $\alpha$, and thus the perturbative expansion can
simply be formulated as those in terms of the bare parameter in the
Dirac operator.
Although the difficulty caused by the severe infrared divergences may
be common, the lattice formulation of the calculation is significantly
simplified by concentrating on that part.

With the understanding of the possible uncertainties of the method
studied in Ref.~\cite{Kitano:2022gzy}, we performed simulations with
large lattice volumes on supercomputers.
We evaluated the perturbative coefficients up to five loops, and
obtained consistent results with those of two groups.

\section{Lattice formulation of quenched QED}

Since we concentrate on the Feynman diagrams without lepton loops, one
can ignore the fermion action in generating the configurations of the
photon field. As it is usual in the lattice simulation, the
correlation functions are obtained as the statistical averages of
quantities evaluated under each configurations.

The QED without the fermion action is a free theory. The Euclidean
lattice action in the Feynman gauge can be written as
\begin{align}
    S_{\rm QED} = {1 \over 2}
    \sum_{n, \mu}
    \hat A_\mu (n) (-\nabla^2 + m_\gamma^2) \hat A_\mu (n).
    \label{eq:action_mod1}
\end{align}
Here we take the unit of $a=1$ with $a$ the lattice spacing. We
recover $a$ in the following discussion when it is necessary.
The photon field, $\hat A_\mu (n)$, is a ultraviolet (UV) regularized
one defined by acting a differential operator,
\begin{align}
     \hat A_\mu (n) \equiv H \left( -{\nabla^2 \over \Lambda_{\rm
UV}^2} \right) A_\mu(n),
\end{align}
with a function $H(x)$ to satisfy $H(x) \to 0$ for $x \gg 1$ and $H(x)
\to 1$ for $x \ll 1$.
In this study, we take a sharp UV cut-off,
\begin{align}
    H(x) = \left \{
        \begin{array}{ll}
            0, & x \ge 1,\\
            1, & x < 1.
        \end{array}
        \right.
\end{align}
In QED, this simple cut-off is gauge invariant.
In the momentum space, the action is diagonalized, and the path
integral is a product of gaussian integrals for each momentum modes.
For momentum modes with $k^2 \ge \Lambda_{\rm UV}^2$, the UV regulator
eliminates their fluctuations while lower modes untouched.
The IR divergence for the $k^2= 0$ mode is regularized by the finite
photon mass, $m_\gamma$.
By this simple set-up, one can generate the photon configurations
trivially by generating gaussian noises for each momentum modes up to
$k^2 < \Lambda_{\rm UV}^2$ with the variance of $1/(k^2 + m_\gamma^2)$
and then performing Fourier transform to get back to the position
space.
The UV cut-off is not formally necessary as the finite lattice spacing
in a finite box already replaces the momentum integral to a finite
summation. Nevertheless, the introduction of $\Lambda_{\rm UV}$ is
numerically necessary in order to suppresses large logarithmic factors
which make the continuum limit further~\cite{Kitano:2021ecc}.

For the calculations of the correlation functions, we use the naive
Dirac operator on the lattice:
\begin{align}
    (D)_{nm}^{\alpha \beta} = m \delta_{nm} \delta_{\alpha \beta}
   + {1 \over 2} \sum_\mu \left[
   (\gamma_\mu)_{\alpha \beta} e^{i e A_\mu (n)} \delta_{n+\mu,m}
   - (\gamma_\mu)_{\alpha \beta} e^{-i e A_\mu (n-\mu)} \delta_{n-\mu,m}
   \right].
\end{align}
There are unwanted doublers in the propagators obtained by inverting
this Dirac operator. We will take care of those unwanted modes later.
The parameter $m$ is the fermion mass and $e$ is the coupling
constant. We formally expand every quantities in terms of $e$, and
evaluate the statistical averages of each coefficients. In this way,
at any stage of the calculation, we do not need a specific value of
$e$. The coupling constant $e$ is not a parameter in the simulation.
This part is different from lattice simulations of non-perturbative
dynamics where the continuum limit is taken by tuning the coupling
constant to a critical value.

Since there is no lepton loop in our study, $e$ will not be
renormalized. This means that $e$ can be identified as the physical
coupling constant, i.e., $\alpha = e^2 / (4 \pi) \simeq 1/137$,
otherwise, we would need to re-expand $e$ by physical $\alpha$. The
perturbative coefficients we evaluate are going to be the physical
quantities after taking various limits discussed below and can be
compared with the ones from the Feynman diagram calculations.

The limits we should take are $\Lambda_{\rm UV} \to \infty$, $m_\gamma
\to 0$, and $a\to 0$ while fixing the fermion mass $m$ to be finite to
bring back to the continuum QED. In practice, we fix $\Lambda_{\rm UV}
a = 1.5$, and take the limit of $m_\gamma / m \to 0$ and $ma \to 0$.
We also need to take $L \to \infty$ and $T \to \infty$, where $L$ and
$T$ are the lattice sizes in the spacial and the temporal directions,
respectively. Since we introduce the photon mass $m_\gamma$ as the IR
regulator, the finite volume effects are suppressed exponentially as
$e^{-m_\gamma L}$ and $e^{-m_\gamma T}$. Therefore, we keep the
combinations of $m_\gamma L$ and $m_\gamma T$ large enough in the
calculation so that we do not have to deal with $O(1/L)$ or $O(1/T)$
corrections.

\section{$g$-factor measurements on the lattice}

We measure fermion-fermion-current three point functions to obtain the
form factors.
We work in the position space in the Euclidean temporal direction and
in the momentum space in the spacial direction.

We evaluate the three-point function as follows. Under each photon
configuration, $A_\mu (n)$, which is generated according to the free
field theory, we evaluate the quantity inside the bracket of
\begin{align}
  G_\mu (t) = \Big \langle \sum_{\bf p'}
  D^{-1} (t_{\rm sink}, t; {\bf p}, {\bf p}') \gamma_\mu
  D^{-1} (t, t_{\rm src}; {\bf p' + k}, {\bf p+k}) \Big \rangle
  ,
\label{eq:three-point}
\end{align}
and take the ensemble average, $\langle \cdots \rangle$.
The locations $t_{\rm src}$, $t_{\rm sink}$ and $t$ are those of two
fermions and the current operator, respectively. We fix $t_{\rm src}$
and $t_{\rm sink}$ at some particular locations, and view the
correlation function as a function of $t$ for later purpose.
The momenta ${\bf p}$ and ${\bf k}$ are those of the outgoing fermion
and the photon (the current vertex). We take ${\bf k}$ to be the
smallest non-zero momentum on the lattice, ${\bf k} = (0,0,2 \pi /
L)$, and ${\bf p} = -{\bf k}/ 2$ so that the incoming and the outgoing
fermion momenta, ${\bf p+k}$ and ${\bf p}$, are in the back-to-back
configuration. The half-integer momentum of the fermions can be
realized by taking the anti-periodic boundary condition in the $z$
direction. For the $g$ factor calculation, we need to take the limit
of ${\bf p} = {\bf k} = 0$. This causes an error of $O(1/L^2)$ that we
ignore.
The inverse of the Dirac operator represents the propagation of the
fermion under the photon background.
In the calculation of the no-lepton-loop diagrams, one can ignore the
disconnected diagram which directly connects $t_{\rm src}$ to $t_{\rm
sink}$ and $t$ to $t$, since the latter propagator forms a fermion
loop.
All the Feynman diagrams are, therefore, included in
Eq.~\eqref{eq:three-point}.
In the definition of the three-point function, we use the naive
current operator $\bar \psi \gamma_\mu \psi$, which is non-conserving
on the lattice. The renormalization of the vertex will be discussed
later.

The inversion of the Dirac operator, $D$, can be taken perturbatively
by the use of the Fourier transform~\cite{DiRenzo:2000qe}.
This is basically the same procedure as the Feynman diagram
calculation. The full propagator of the fermion at each order in the
perturbative expansion is given by the product of the free propagators
and the vertices. The vertices are associated with the photon field
$A_\mu (n)$ that is a function in the position space. The free fermion
propagator is a diagonal matrix (a function of a single momentum
$k_\mu$) in the momentum space. The multiplication of functions, i.e.,
diagonal matrices, can be done quickly on computers, and the Fourier
transform between the position and momentum spaces can also be
effectively performed by the use of the fast Fourier transform (FFT)
algorithm.
In this way, the multiple-dimensional integrations of loop momenta in
the Feynman diagram calculation are numerically encoded as a series of
FFTs and all the different Feynman diagrams are summarized as a single
term in Eq.~\eqref{eq:three-point}.

For large enough separations of $t_{\rm sink}$, $t_{\rm src}$, and
$t$, the three-point function is dominated by contributions from the
lowest energy state, that is the on-shell fermion. One can extract the
physical form factors of the fermion by taking such a set-up.
One complication is that there are also contributions from the doubler
state in the time direction. The doubler contribution gives
alternative signs for even and odd values of $t_{\rm sink} - t_{\rm
src}$ because the on-shell pole is at $E = m_* + i \pi$ and the
propagator is proportional to $e^{- E t}$. One can therefore remove
them by taking an average with the next site.
For the choice of our external momentum, ${\bf p} = -({\bf p + k})$,
an easier treatment is possible thanks to the parity symmetry of the
fermion configuration. The doubler contributions to the form factors
are identical to the physical ones for odd separations of $t_{\rm
sink} - t_{\rm src}$, whereas it completely cancels out the physical
ones for even separations.
Therefore, one can remove the doubler contributions by setting an odd
separation of $t_{\rm sink} - t_{\rm src}$ and taking an appropriate
normalization as we see later.
For spacial directions, there is no doubler contribution since we work
in the momentum space.

Here we define projections to the electric and the magnetic functions:
\begin{align}
    G_E (t) = {\rm tr} \left[
        {1 + \gamma_4 \over 2} G_4 (t)
    \right], \quad
    G_M (t) = i \sum_{i,j,k} \epsilon_{ijk} {\rm tr} \left[
        {1 + \gamma_4 \over 2} 
        \gamma_5 \gamma_i G_j (t)
    \right] {\bf k}_k,
\end{align}
where $i,j,k$ are indices for the spacial directions, $x$, $y$ and
$z$. The Euclidean temporal direction is labelled as 4. The trace is
taken over the spinor indices. The Dirac representation of the gamma
matrices is convenient here since the projection, $(1+\gamma_4)/2$,
simply picks up the upper two components of the source and sink
spinors.
One can take away the external fermion legs and obtain the form
factors by
\begin{align}
    F_E (t) = {G_E(t) \over G_E^{\rm norm} (t)}, \quad
    F_M (t) = {G_M(t) \over G_M^{\rm norm} (t)},
    \label{eq:ratio}
\end{align}
where $G_E^{\rm norm} (t)$ and $G_M^{\rm norm} (t)$ are calculated
from the combination of the two-point functions:
\begin{align}
    G_\mu^{\rm norm} (t) = \sum_{\bf p'}
    \Big \langle D^{-1} (t_{\rm sink}, t; {\bf p}, {\bf p}') \Big \rangle
    \gamma_\mu
    \Big \langle D^{-1} (t, t_{\rm src}; {\bf p' + k}, {\bf p+k}) \Big \rangle
    .
    \label{eq:norm}
\end{align}
The doubler contribution is cancelled as both numerators and
denominators are multiplied by two.
The $g$ factor can now be obtained as the ratio of the electric and
the magnetic form factors:
\begin{align}
    {g (t) \over 2} = {F_M (t) \over F_E (t)},
    \label{eq:g-factor}
\end{align}
up to a correction from finite ${\bf k}$. For large enough separations
of $t$ from $t_{\rm src}$ and $t_{\rm sink}$, $g(t)$ should be
independent of $t$.
By looking for a plateau as a function of $t$, one can obtain the $g$
factor numerically.
Again, all the quantities are expanded in terms of the power series of
$e$ at any stage of calculation. Namely, the multiplication and the
inversion are encoded as the rules for each perturbative coefficients
in the computer program. Therefore, at the end of the calculation, we
obtain numerical evaluation of the coefficients.

The renormalization is implicitly done in this flow of calculation.
For taking away the fermion legs, we use the full propagator in
Eq.~\eqref{eq:norm}. This can be thought of as the wave function and
the mass renormalizations.
In Eq.~\eqref{eq:g-factor}, we take the ratio of the electric and the
magnetic form factors.
This is nothing but the vertex renormalization to normalize $F_E$ to
unity.

The method described here is slightly different from the previous
works in Refs.~\cite{Kitano:2021ecc, Kitano:2022gzy}, where the
correlation functions are calculated in the momentum space and the
double pole part associated with two fermion propagators are extracted
by the Fourier transform and extrapolation.
Two methods give the same form factors in the continuum limit, but we
take the position space calculation here since it is more transparent.

\section{Numerical results}

We perform the lattice simulations with five sets of lattice volumes,
$L^3 \times T$ = $24^3 \times 48$, $28^3 \times 56$, $32^3 \times 64$,
$48^3 \times 96$, and $64^3 \times 128$, where simulations are
performed on NEC SX-Aurora TSUBASA A500-64 at KEK for smaller four
volumes up to $48^3 \times 96$, and on supercomputer Fugaku at RIKEN
for $64^3 \times 128$. We also performed a simulation with the $96^3
\times 192$ lattice volume on Fugaku with small statistics as a trial.
For the Fugaku platform, we developed a computer code to make high
parallelization possible based on the public lattice simulation code
\bridge~\cite{Ueda:2014rya}.

For each lattice volume, we fix the fermion mass parameter to be $ma =
2.04/\sqrt L$, and take five points of the photon mass parameter,
$m_\gamma a = 4/L$, $5/L$, $6/L$, $7/L$, and $8/L$. (We take only
$m_\gamma a = 4/L$ for $L=96$.) The UV cut-off is fixed as
$\Lambda_{\rm UV} a = 1.5$ for all the volumes.
The source location is fixed as $t_{\rm src} = 0$ and $t_{\rm sink}$
is taken to be the largest odd integer below $3T/4$.
We take the average of the periodic and anti-periodic boundary
conditions in the temporal direction in evaluating the inverse of the
Dirac operators~\cite{Kitano:2022gzy}. This treatment effectively
doubles the time extent $T$. The contribution from the backward
propagation on the torus to connect $t_{\rm src}$ and $t_{\rm sink} -
T \sim t_{\rm sink}$ is cancelled even though it is actually closer
for our choice of $t_{\rm src}$ and $t_{\rm sink}$.

The finite volume effects are controlled by the $e^{-m_\gamma L}$
factor, that is at most a few percent for each parameter points.
Barring this level of errors, one can interpret all the data to be
equivalent to the ones obtained with infinite volume with finite
$m_\gamma$ and $a$.
For larger $L$ simulations, one can take smaller values of $m_\gamma /
m$ as well as smaller $ma$. As explained before, we need to take two
limits $ma \to 0$ and $m_\gamma / m \to 0$ while $\Lambda_{\rm UV} a$
fixed to reach to the continuum QED.

\begin{figure}[p]
    \includegraphics[width=0.43\linewidth]{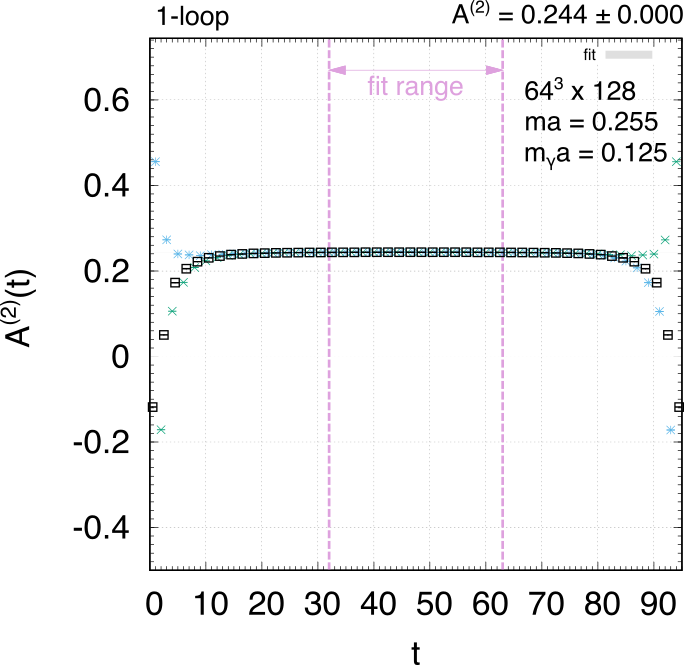}\hspace{0.5cm}
    \includegraphics[width=0.43\linewidth]{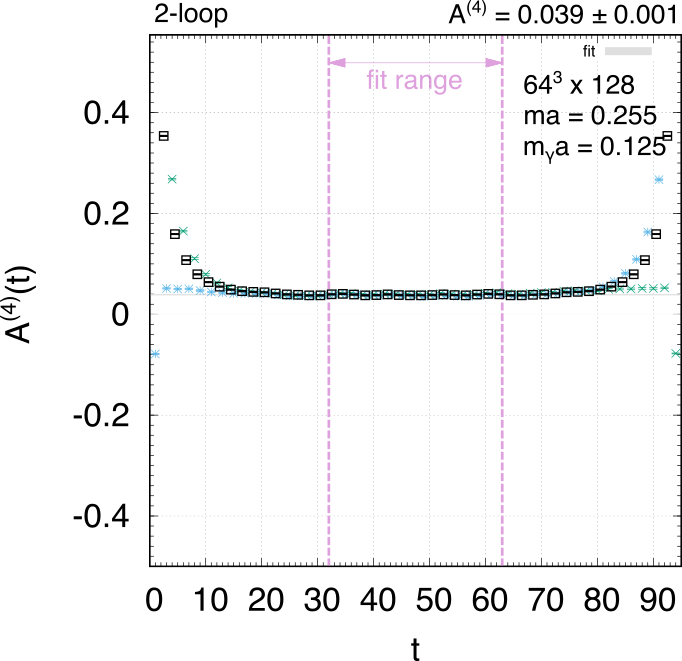}\\ \\
    \includegraphics[width=0.43\linewidth]{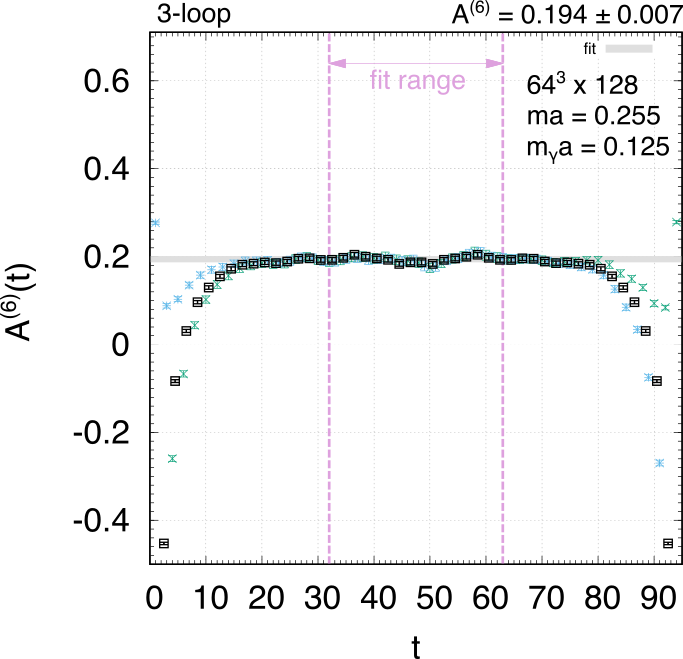}\hspace{0.5cm}
    \includegraphics[width=0.43\linewidth]{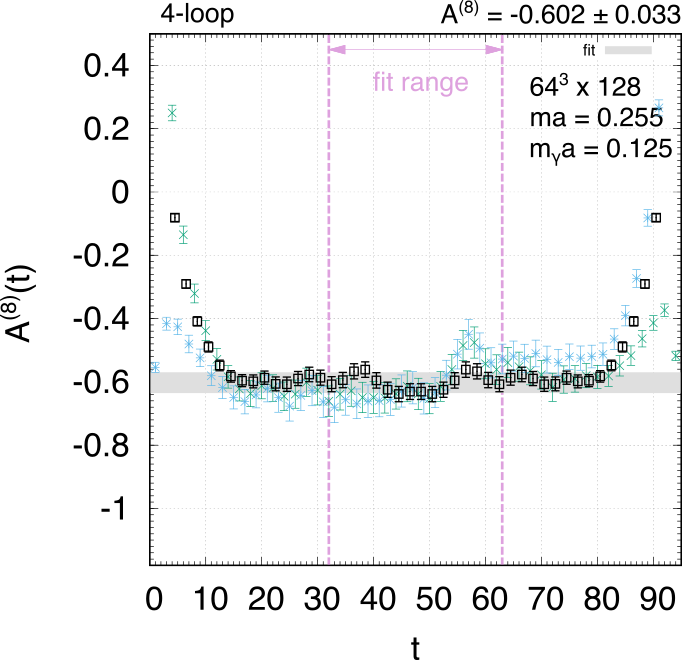}\\ \\
    \includegraphics[width=0.43\linewidth]{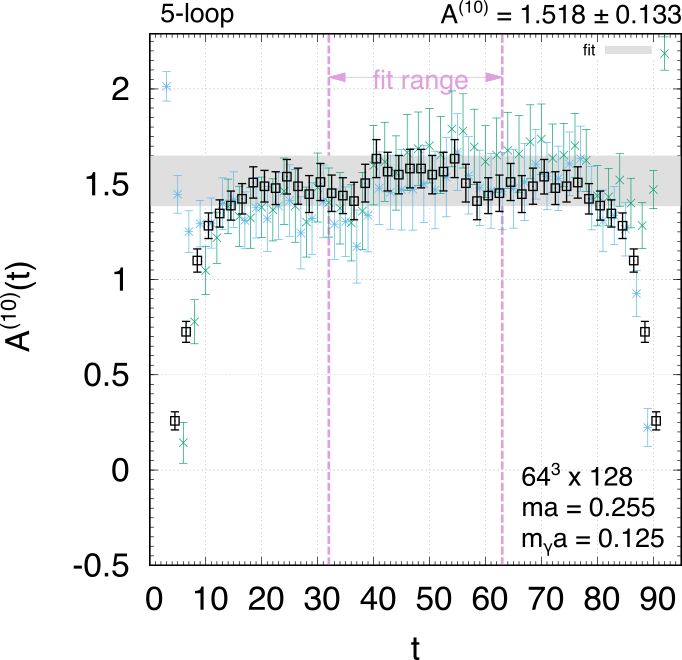}
\caption{The perturbative coefficients of $g(t)/2$ defined in
Eq.~\eqref{eq:g-factor} and \eqref{eq:expand} for the $64^3 \times
128$ lattice with $m_\gamma a = 0.125$ and $ma=0.255$. The left
and right end of $t$ is the source and the sink location, and $t$
is the location of the current operator.}
\label{fig:gt}
\end{figure}

We show in Fig.~\ref{fig:gt} the perturbative coefficients
$A^{(2n)}(t)$ of the $g(t)$ function in Eq.~\eqref{eq:g-factor}
defined as
\begin{align}
    {g(t) \over 2} = 1 
    + A^{(2)} (t) \left( {\alpha \over \pi} \right)  
    + A^{(4)} (t) \left( {\alpha \over \pi} \right)^2 + \cdots.
    \label{eq:expand}
\end{align}
We show data for the $64^3 \times 128$ lattice with $m_\gamma a = 8/L
= 0.125$. We analyzed about 200,000 configurations for this parameter
set. One can observe plateaus in the middle region of $t$. The data
for even (green $\times \llap +$) and odd $t$ (blue $\times$) behave
differently due to the contributions from doublers associated with the
higher excited modes. The average between next two points are shown as
black squares.
We take the average of the middle region ($T/4 \leq t < T/2$) as the
value of the plateau, and define the average as $A^{(2n)}$ at the
$n$-loop order.

\begin{figure}[p]
    \includegraphics[width=0.43\linewidth]{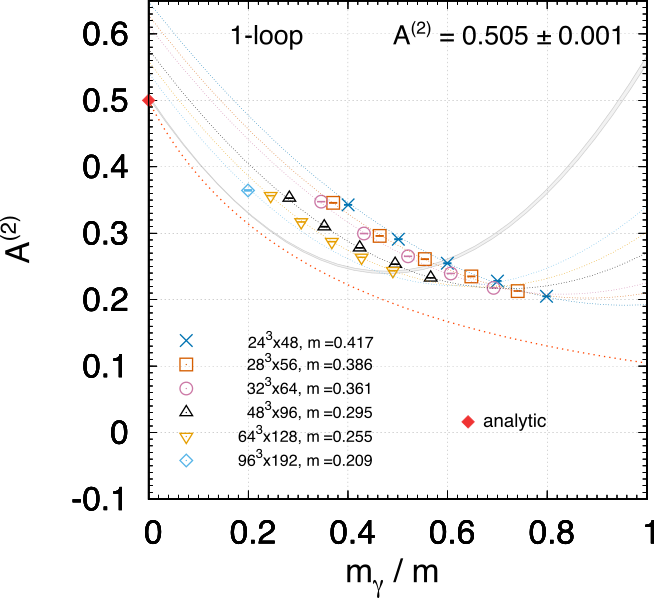}\hspace{0.5cm}
    \includegraphics[width=0.43\linewidth]{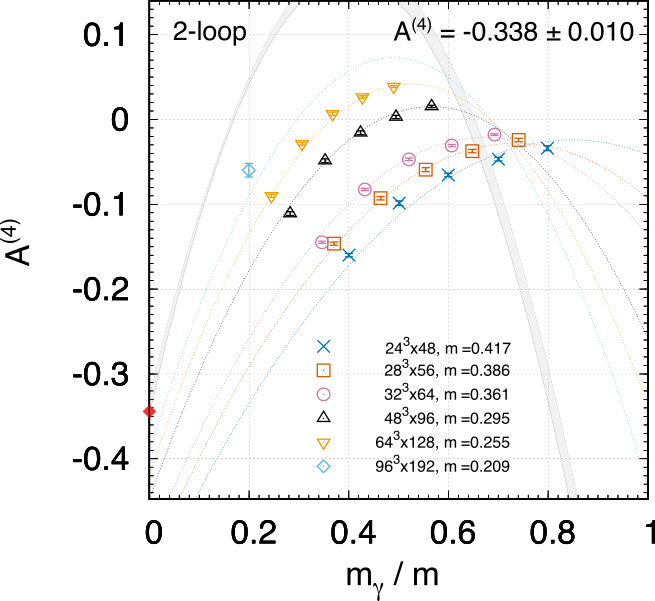}\\ \\
    \includegraphics[width=0.43\linewidth]{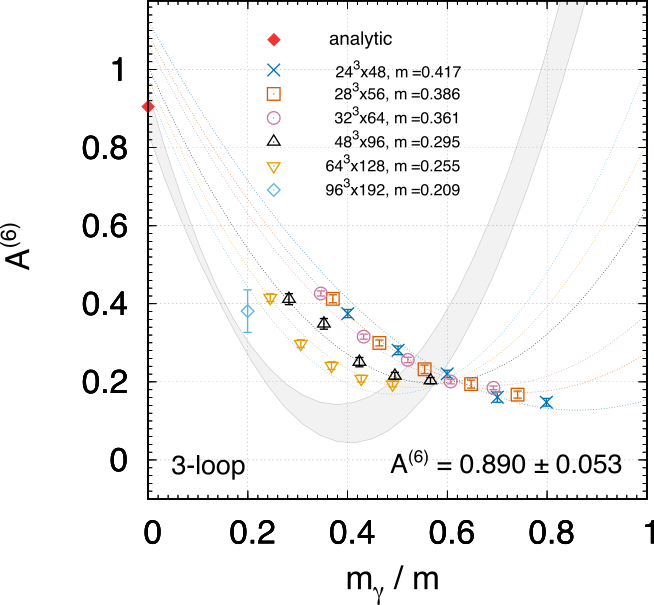}\hspace{0.5cm}
    \includegraphics[width=0.43\linewidth]{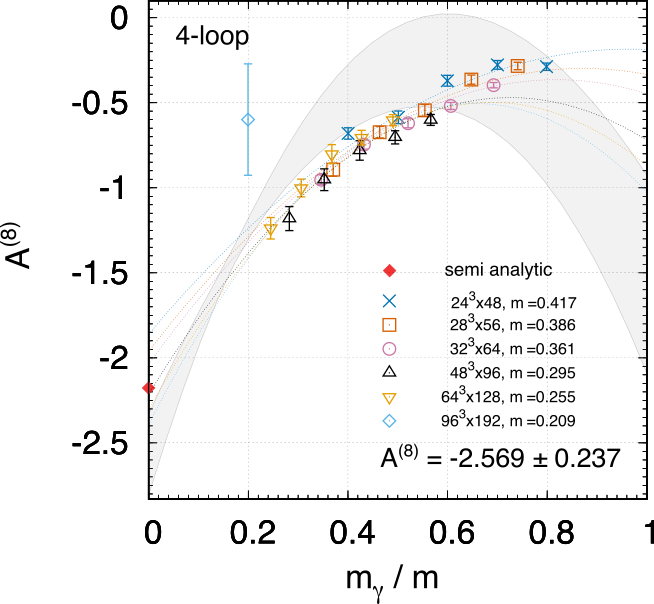}\\ \\
    \includegraphics[width=0.43\linewidth]{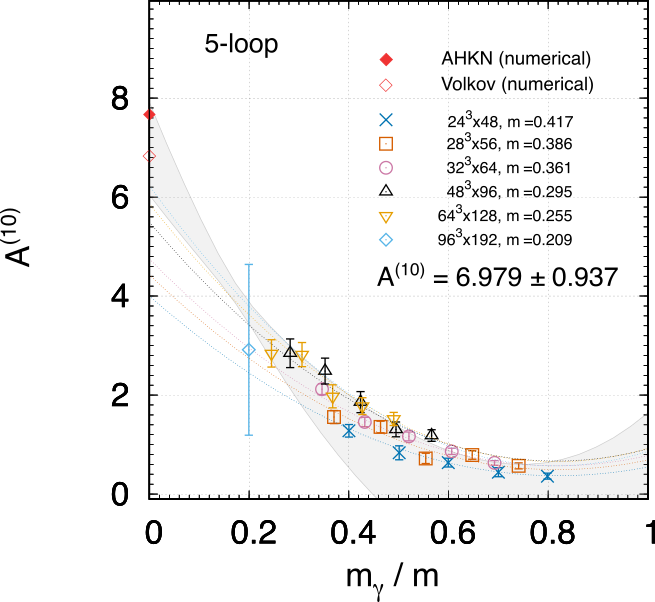}
\caption{The coefficients of the perturbative expansion of $g/2$ at
finite $m_\gamma a $ and $ma$. By the fitting with
Eq.~\eqref{eq:fitting}, we obtain the continuum limit as the gray
band. The red diamonds represent the known results by the Feynman
diagram method to be compared with the section of the gray band at
$m_\gamma / m = 0$. The red dotted curve at the one-loop level is the
analytic results in the continuum theory.}
\label{fig:g}
\end{figure}

The values of $A^{(2n)}$ for each parameter set are shown as a function of
$m_\gamma / m$ in Fig.~\ref{fig:g}.
We perform fitting of those points by the following function to
extrapolate to $ma \to 0$ and $m_\gamma / m \to 0$:
\begin{align}
    A^{(2n)} (m, m_\gamma)
    =& a_0^{(2n)} \left(1 + b_0^{(2n)} (ma)^2 \right) \nonumber \\
    &+ a_1^{(2n)} {m_\gamma \over m} \left(1 + b_1^{(2n)} (ma) \right)
    + a_2^{(2n)}  \left( {m_\gamma \over m} \right)^2 
    \left(1 + b_2^{(2n)} (ma) \right),
    \label{eq:fitting}
\end{align}
with $a_{0-2}^{(2n)}$ and $b_{0-2}^{(2n)}$ as parameters for each
$n$-loop order. The results of the fitting to the continuum limit, $ma
\to 0$, are shown as gray bands (1$\sigma$) in each figure.
We take into account up to the quadratic order in $m_\gamma / m$, and
the leading correction in terms of $ma$ is included. The leading
correction at the zero-th order of $m_\gamma / m$ is $O((ma)^2)$ due
to chiral symmetry. Since we maintain chiral symmetry on the lattice
by using the naive Dirac operator, there is no $O(ma)$ correction
towards the continuum limit.
For higher orders in $m_\gamma / m$, there can be $O(m_\gamma a)$ or
$O(m_\gamma^2 a / m)$ corrections, that we take into account in the
fitting.
As one can see from the figures, the $m_\gamma / m \to 0$ limit is
quite far from the data points we obtained. We cannot, however, take
arbitrarily small $m_\gamma / m$ as we need to keep the combination of
$m_\gamma L$ to be large enough to suppress the finite volume
corrections while $ma$ needs to be close enough to the continuum
limit, $ma \to 0$.
This difficulty can be understood as reflection of severe IR
divergences in this set of diagrams.

At one-loop, the red dotted curve is the analytic calculation of
$A^{(2)}$ as a function of the finite photon mass in the continuum
theory. 
Even though it is a quite large extrapolation, we find a good
agreement at the small $m_\gamma / m$ region.
The fitting results at $m_\gamma /m = 0$ at each order should be
compared with the red diamonds which represent the known results from
the diagrammatic calculations. (We take the values up to four loops
from the table in Ref.~\cite{Volkov:2017xaq}. For five loops, see
below for explanation.) The fitting results at each order are shown in
the figures.
We see quite good agreements. The results seem to be indicating that
the systematic uncertainties, such as the finite volume effects and
the one associated with the fitting function are under good control.

Our estimation of the five-loop coefficients, $A^{(10)}$, is  
\begin{align}
    A^{(10)} ({\mbox{no lepton loop}}) = 7.0 \pm 0.9,
\end{align}
where only the statistical error is taken into account. The value is
consistent with the result of Ref.~\cite{Volkov:2024yzc} ($6.828 \pm
0.060$, open diamond). There was a tension between this value and the
AHKN result ($7.668 \pm 0.159$, filled diamond) in
Refs.~\cite{Aoyama:2012wk, Aoyama:2019ryr}\footnote{The value is
obtained by the subtraction of the with-lepton-loop part in
Ref.~\cite{Aoyama:2012wk} from the total value in
Ref.~\cite{Aoyama:2019ryr}. 
The obtained value is consistent with the one given in
Ref.~\cite{Aoyama:2017uqe}.
See Ref.~\cite{Volkov:2024yzc} for a more
detail explanation.}. It has been reported recently that a new
preliminary AHKN result is now consistent with
Ref.~\cite{Volkov:2024yzc} within
1$\sigma$~\cite{g-2_theory_initiative2024}.
In any case, our result seems to be giving a good estimate of the
five-loop coefficient.

\section{Summary}

We continued the trial to obtain the perturbative coefficients of
$g-2$ in QED by using lattice simulations.
The calculation is quite simple as we only have a single diagram to
evaluate.
By concentrating on the diagrams without a lepton loop, the
formulation is further simplified as one can generate photon
configurations by the free theory.
We obtain results which are consistent with the Feynman diagram
method.
Further improvement of the statistical precision seems to be possible.
The data points of the largest lattice, $96^3 \times 192$, are
obtained with a trial run with about 20,000 configurations. Ten times
large statistics with several different values of $m_\gamma$ should
easily be possible if computational time is allocated.

Although one probably needs to be more careful about the possible
systematic errors to give more reliable predictions, our calculation
here is meant to be the confirmation of the previous calculations by
estimating the size of the perturbative coefficients by a totally
independent method.
If we need to estimate the coefficient at the six-loop order by future
progress in experimental precisions (which is indeed
expected~\cite{Fan:2022eto}), the method presented here may serve as
the first thing to try as the computational cost scales only as
$(2n)^2$ with $n$ the loop order while the number of Feynman diagrams
grows as more than $(2n)!$~\cite{PhysRev.91.1243, Cvitanovic:1978wc}.

Including the with-lepton-loop part is possible along the line of the
method in Ref.~\cite{Kitano:2021ecc}.
In this case, we need to generate configurations according to the
Langevin equations~\cite{DiRenzo:2000qe} where the inversion of the
Dirac operator is needed in each Langevin step.
Although the formulation is more complicated, the cost of the
computation may not be so heavy since less severe IR divergence may
mean that we need less statistics.

\section*{Acknowledgements}
We thank Hideo Matsufuru for his support for developing the computer
code based on
\bridge~(\href{http://bridge.kek.jp/Lattice-code/}{http://bridge.kek.jp/Lattice-code/}).
This work is supported in part by JSPS KAKENHI Grant-in-Aid for
Scientific Research (Nos.~19H00689, 21H01086, 23K20847, and 22K21350).
This work used computational resources of supercomputer Fugaku
provided by the RIKEN Center for Computational Science through the
HPCI System Research Project (Project ID: hp230463, hp230511).
This work is also supported by the Particle, Nuclear and Astro Physics
Simulation Program No.~007 (FY2019), No.~001 (FY2020), No.~003
(FY2021), and No.~001 (FY2022) of Institute of Particle and Nuclear
Studies, High Energy Accelerator Research Organization (KEK).

\appendix

\bibliographystyle{./utphys.bst}
\bibliography{./g-2.bib}

\end{document}